\documentclass[USenglish,twocolumn]{article}

\ifx\directlua\undefined\ifx\XeTeXcharclass\undefined
  \usepackage[utf8]{inputenc}                           %pdftex engine
  \else\RequirePackage[no-math]{fontspec}[2017/03/31]\fi %xetex engine
  \else\RequirePackage[no-math]{fontspec}[2017/03/31]\fi %luatex engine
\usepackage[sort&compress,square,numbers]{natbib}
\usepackage[big,online]{dgruyter}
\usepackage{xcolor}

\theoremstyle{dgthm}

\theoremstyle{dgdef}

\begin{document}

\articletype{Research Article}

\author[1]{Likai Yang}
\author[2]{Jiacheng Xie}
\author*[3]{Hong X. Tang}
\affil[1]{Department of Electrical Engineering, Yale University, New Haven, CT 06511, USA}
\affil[2]{Department of Electrical Engineering, Yale University, New Haven, CT 06511, USA}
\affil[3]{Department of Electrical Engineering, Yale University, New Haven, CT 06511, USA, email: hong.tang@yale.edu}
\title{Fluorescence enabled phonon counting in an erbium doped piezo-optomechanical microcavity}
\runningtitle{}
\abstract{Converting phonons to photons with optomechanical interaction provides a pathway to realize single phonon counting, which is instrumental in the quantum applications of mechanical systems such as entanglement generation, thermometry, and study of macroscopic quantum phenomenon. In this process, the key requirement is high-extinction, narrow-bandwidth, and stable filtering of the parametric optical pump. Here, we propose to lift this necessity by counting fluorescence emission from a rare earth embedded optomechanical cavity. By doing so, we show that an equivalent filtering effect can be achieved due to spectral hole burning and cavity Purcell effect. To demonstrate this, we designed, fabricated, and characterized an integrated piezo-optomechanical Fabry-Perot cavity on the erbium doped thin-film lithium niobate platform. By collecting fluorescence from the optomechanical sideband, we show that 93\,dB suppression of the pump can be achieved with 10\,dB loss of signal, resulting in an increase of 83\,dB in sideband-pump ratio. Our results facilitate a route to realize filterless single phonon counting and also create new opportunities to study the interaction between solid state emitters and mechanical systems.}
\journalname{Nanophotonics}
\journalyear{2024}
\journalvolume{aop}

\maketitle

\section{Introduction} 

Investigating micromechanical systems within the quantum regime opens up new avenues for understanding macroscopic quantum effects, and for various applications in quantum metrology and information processing. For example, cooling of mechanical resonators to quantum ground states has been demonstrated \cite{rocheleau2010preparation}. Coupling between acoustic resonators and qubit systems such as superconducting qubits \cite{chu2017quantum} and NV centers \cite{macquarrie2013mechanical} has been achieved for efficient qubit control. Acting as an intermediate system to bridge microwave and optical photons, piezo-optomechanical devices have also been exploited to realize efficient quantum transducers \cite{jiang2020efficient}. These studies on quantum acoustics can significantly benefit from the ability to detect single quanta of mechanical motion, i.e. phonons. As an analogy, the presence of efficient single photon detectors in the optical domain has enable a wide range of quantum applications including high-sensitivity imaging \cite{seitz2011single}, entanglement generation \cite{wang2016experimental}, and mapping photon statistics \cite{hlouvsek2019accurate}. While similar applications can be envisioned with mechanical systems, direct counting of single phonons can be challenging. Alternatively, it can be realized by converting phonons to photons via optomechanical scattering. Detection of scattered photons is then equivalent to counting the phonons. Using this method, experiments such as ground state thermometry \cite{meenehan2015pulsed} and phonon addition \cite{enzian2021single} have been realized by extracting the phonon occupancy of mechanical resonators. Phonon intensity interferometry \cite{cohen2015phonon} and non-classical phonon-photon correlation \cite{riedinger2016non} have also been demonstrated thanks to the efficient detection of single phonons. 

\begin{figure}[t]
\includegraphics[width=0.5\textwidth]{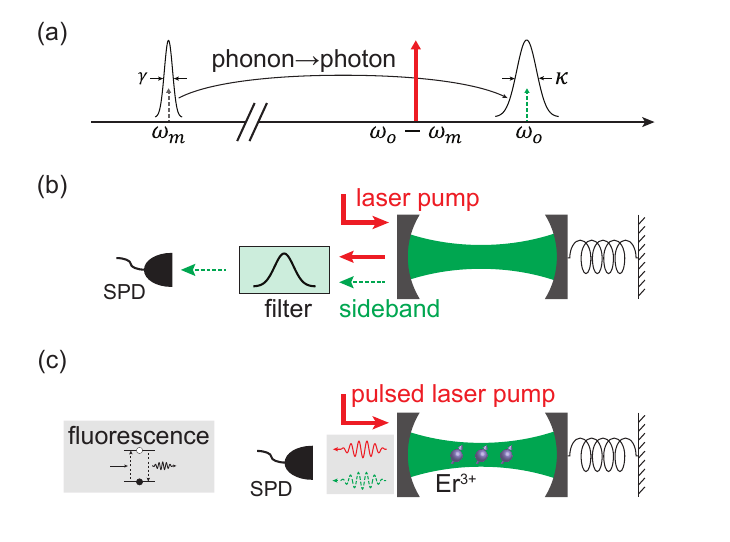}% Here is how to import EPS art
\caption{(a) Schematic drawing of the optomechanical conversion process. A mechanical mode at $\omega_m$ with damping rate $\gamma$ and an optical mode at $\omega_o$ with damping rate $\kappa$ is co-localized to established the coupling. The phonon-to-photon conversion can be realized by introducing a red-detuned pump at $\omega_p=\omega_o-\omega_m$. (b) Direct counting of sideband photons, in which high-extinction and narrow-band filtering of the pump is required. The dashed line used for the sideband indicates that it is usually much weaker than the pump and at single photon level. SPD: single photon detector. (c) By doping rare earth ions in the cavity, the fluorescence emission following a pulsed pump can be counted with equivalent filtering effect from spectral hole burning and cavity Purcell enhancement.}
\label{fig1}
\end{figure}

In the phonon counting experiments, optical ($\omega_o$) and mechanical ($\omega_m$) resonance modes are typically used and co-localized to enhance the coupling. The process of phonon-to-photon conversion is illustrated in Fig.~\ref{fig1}a. By applying a red-detuned optical pump at $\omega_p=\omega_o-\omega_m$, parametric coupling between optical and mechanical modes is established. This pump frequency is chosen so that the sideband frequency $\omega_p+\omega_m$ aligns with the optical resonance, thus maximizing the sideband intensity. The effective Hamiltonian of the sytem can be denoted by $\hat{H}_{\mathrm{int}}=\hbar G(\hat{a}_m^\dagger\hat{a}_o+\hat{a}_o^\dagger\hat{a}_m)$. Here, $G=\sqrt{n_o}g_0$ is the photon number enhanced coupling strength, with $n_o$ being the cavity photon number introduced by the detuned pump, and $g_0$ being the single-photon optomechanical coupling strength. To explicitly count the converted photons in the sideband, the strong pump light needs to be suppressed with high-extinction and narrow-band filtering, as sketched in Fig.~\ref{fig1}b. The performance of these external filters, typically composed of cascaded Fabry-Perot cavities that require active stabilization \cite{galinskiy2020phonon}, often becomes the limiting factor in determining the measurement sensitivity \cite{cohen2015phonon,meenehan2015pulsed}.

In this work, we propose a novel method to realize sideband phonon counting using fluorescence emission from a rare earth doped piezo-optomechanical cavity. Thanks to their ultra-narrow optical linewidth, i.e. long coherence, rare earth ions (REIs) have found numerous applications in quantum information processing including quantum memory \cite{tittel2010photon,zhong2017nanophotonic}, transducer \cite{williamson2014magneto,bartholomew2020chip}, and single photon source \cite{dibos2018atomic}. For the same reason, REI doped materials have also been explored as narrow-band filters utilizing the spectral hole burning technique \cite{thiel2011rare,beavan2013demonstration,ulrich2022high,zhao2024cavity}, in which narrow spectral features can be created within the inhomogeneous profile by saturating the ions at certain frequencies. The linewidth of the spectral hole, thus the filter bandwidth, can be as narrow as the homogeneous linewidth, which can readily reach kHz level at cryogenic temperature of several Kelvins \cite{thiel2010optical}. In recent years, there have been extensive interests in coupling REIs with on-chip integrated photonic cavities to modify their emission properties \cite{dibos2018atomic,zhong2018optically,yang2023controlling}. Exhibiting high quality factors and small mode volume, these cavities can greatly enhance the emission rate from REIs via the Purcell effect \cite{merkel2020coherent}.

In our proposed scheme, a pulsed drive is sent to a REI doped optomechanical cavity (Fig.~\ref{fig1}c). Ions at pump ($\omega_p=\omega_o-\omega_m$) and sideband ($\omega_o$) frequency will be excited and subsequently fluorescent at their respective frequencies. Instead of directly counting the sideband signal, the fluorescence emission is measured. Due to the spectral hole burning effect by the strong pump, the REI emission at pump frequency will be saturated, resulting in a much lower excitation-to-fluorescence conversion efficiency. On the other hand, the ions at sideband frequency will remain unaffected by the pump if the mechanical frequency is much larger than the hole linewidth. In other words, most photons from the weak sideband yet only a small percentage of pump photons will be converted to fluorescence, resulting in an equivalent filtering effect. Furthermore, since only the sideband fluorescence is on-resonance with the cavity, it will be amplified by the cavity Purcell enhancement of emission rate. These combined factors would enable the potential to realize optomechanical phonon counting without external filtering.

\begin{figure*}[t]
\centering
\includegraphics[width=0.95\textwidth]{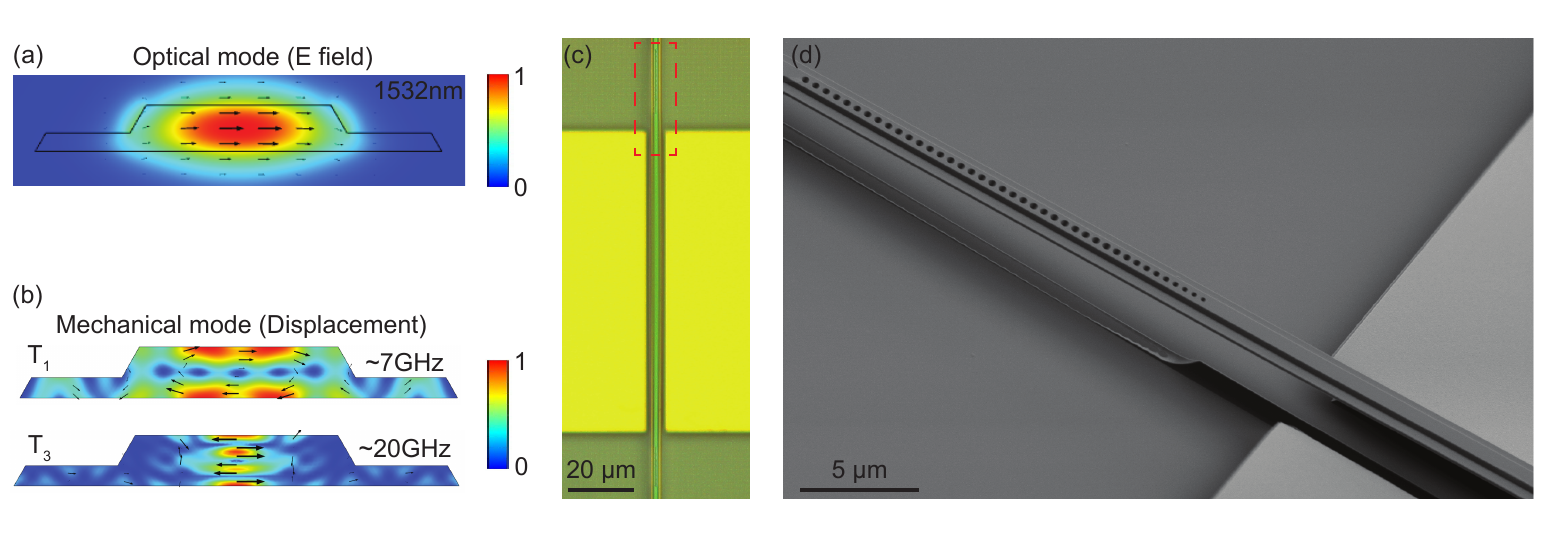}% Here is how to import EPS art
\caption{(a) Simulated fundamental TE optical mode of the waveguide at Er transition wavelength around 1532\,nm. The color bar corresponds to normalized electric field intensity. (b) Cross sectional simulation of the thickness mechanical modes. The first-order ($T_1$) mode is used in the following measurements and the third-order ($T_3$) mode at higher frequency is also shown. The color bar corresponds to normalized mechanical displacement. (c) Optical image of the fabricated device. The device consists of a suspended straight waveguide with two photonic crystal mirrors at each sides, as well as gold electrodes next to the cavity for piezo-electric excitation of mechanical modes. (d) Zoom-in SEM image of the red block area in (c).}
\label{fig2}
\end{figure*}

To demonstrate this effect, we developed a piezo-optomechanical Fabry-Perot microresonator on the smart-cut erbium (Er) doped thin-film lithium niobate (LN) platform \cite{wang2022er}. The favorable piezoelectric properties of LN presents a convenient way to introduce intracavity phonons with electrical drive, thus motivated various studies on electro-mechanical \cite{xie2023sub} and optomechanical \cite{jiang2019lithium} devices. Er ions are chosen as dopants for their widely-interested optical transition around 1550\,nm, within the telecom band. In the following sections, we first introduce the design and fabrication of our device, followed by the characterization of its piezo-optomechanical properties. A microwave-to-mechanical conversion efficiency of 10\,\% and a single photon optomechanical coupling rate $g_0=2\pi\times3.2$\,kHz are extracted. We then proceed to demonstrate the fluorescence phonon counting process by varying the intracavity phonon number with different RF power. By comparing with direct counting, we show that over 93\,dB of pump suppression is achieved with our approach, with around 10\,dB loss of the sideband photons. This gives a 83\,dB improvement in counting sensitivity. These results provide promising aspects toward filterless detection of single phonons with REI fluorescence.

\section{Device}
Our device is designed on a z-cut lithium niobate thin film with 100\,ppm Er doping, prepared by the smart-cut process (NanoLN). The film consists of 300\,nm-thick LN on 2\,\textmu m silicon dioxide (SiO$_2$) substrate on silicon handle. To support both optical and mechanical modes, a straightforward configuration is to design a suspended straight waveguide. The cross section of the waveguide can be simulated with a FEM solver (COMSOL) to have a fundamental TE optical mode, as shown in Fig.~\ref{fig2}a. By removing the SiO$_2$ substrate beneath the waveguide, the waveguide can support in-plane shear mechanical modes confined by its thickness (Fig.~\ref{fig2}b). The first-order mechanical mode ($T_1$) has a simulated frequency around 7\,GHz for 300\,nm thick waveguide. A favorable property of this mode group is the frequency scalability, where higher order mechanical modes can be identified such as the $T_3$ mode at around 20\,GHz shown in the same figure. The high-frequency mechanical modes can be beneficial in future study of quantum acoustics at higher temperature due to less thermal phonon occupancy. To define optical resonance modes, arrays of periodic holes acting as photonic crystal mirrors are patterned at both ends of the waveguide, thus forming a Fabry-Perot optical cavity. By doing so, the spatial overlap between optical and mechanical modes is maximized, while their frequency and free spectral range (FSR) can be conveniently engineered by varying the length and thickness of the waveguide.

The fabrication of the devices are realized with multi-step electron beam lithography (EBL). Half-etch slab waveguides are first patterned using hydrogen silsesquioxane (HSQ) as mask followed by argon (Ar) plasma etching of LN, resulting in a sidewall angle around 60\,$^{\circ}$. The width of the waveguide is set to be 1.2\,{\textmu}m while the etch depth is 180\,nm. The half-etch structure is chosen to reduce the sidewall roughness seen by the optical mode, thus preserving the Q. The photonic crystal holes are then created in the center of the waveguide by fully etching the LN with similar procedure, forming a Fabry-Perot cavity of 100\,{\textmu}m length. The elliptical holes are designed to be 600\,nm$\times$350\,nm, with a lattice constant of 500\,nm. The detailed process to fabricate such photonic crystal structures can be found in our previous work \cite{yang2023controlling}. After that, gold pads with a gap of 6\,{\textmu}m are patterned next to the waveguide using a lift-off process with PMMA resist. By applying voltage to these on-chip electrodes, an in-plane electric field can be generated, thus excites the thickness mechanical modes via the largest piezoelectric coefficient $e_{15}=3.75$\,C/m$^2$ of LN \cite{bouchy2022characterization}. Finally, the chip is dipped into buffered oxide etch (BOE) to release the SiO$_2$ substrate and dried in a critical point dryer (CPD) to form the suspended beam. An optical microscope image of the fabricated device is shown in Fig.~\ref{fig2}c, with a zoom-in SEM image of the mirror region shown in Fig.~\ref{fig2}d. The photonic crystal structure is tapered on the cavity side to increase the reflectivity. One of the two mirrors are also designed to be partially reflective so that the resonance can be accessed by measuring the cavity reflection. An apodized grating coupler is used for fiber-to-waveguide coupling with single-pass efficiency around 20\,\%.

After fabrication, a device with optical resonance at 1532.06\,nm was selected so that the resonance wavelength is within the inhomogeneous broadening of Er transition, with the ground and excited state being the lowest level ($Z_1$) of $^4I_{15/2}$ state and lowest level ($Y_1$) of $^4I_{13/2}$ state, respectively. The device was packaged with fiber glue and wire bonding for optical and microwave access, respectively. It was then cooled down in a dilution refrigerator to 10\,mK for further measurements.

\section{Results} 
\begin{figure}[t]
\includegraphics[width=0.5\textwidth]{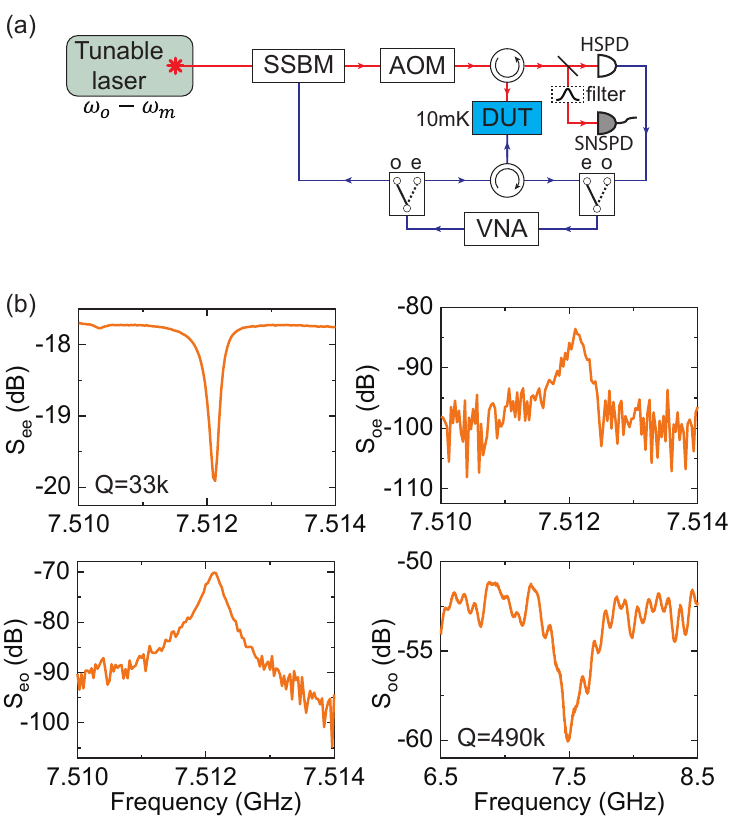}% Here is how to import EPS art
\caption{(a) Schematics of our experimental setup. It consists of two parts, 4-port $S$-parameter measurement and single photon counting. SSBM: single-sideband modulator, AOM: acousto-optic modulator, DUT: device under test, HSPD: high-speed photodetector, SNSPD: superconducting nanowire single photon detector, VNA: vector network analyzer. (b) Spectra of microwave reflection $S_\mathrm{ee}$, microwave-to-optical conversion $S_\mathrm{oe}$, optical-to-microwave conversion $S_\mathrm{eo}$, and optical reflection $S_\mathrm{oo}$. An on-chip pump power of -12\,dBm is used in the measurements.}
\label{fig3}
\end{figure}

The measurement setup used in our experiments is illustrated in Fig.~\ref{fig3}a. Light from a tunable laser (Santec 710) is sent to the device as a parametric pump. Collected via an optical circulator, the response from the device is either demodulated by a high-speed photodetector (HSPD) for $S$-parameter measurements, or counted by a superconducting nanowire single photon detector (SNSPD). The SNSPD used in the experiment (Quantum Opus) has a characterized efficiency of around 50\,\% and a dark count rate around 50\,Hz.

To characterize the optical, mechanical, and conversion properties of our device, a 4-port $S$-parameter calibration \cite{xu2021bidirectional} is first implemented. To do so, the tunable laser is set to red-detuned frequency of $\omega_o-\omega_m$. An RF signal from a vector network analyzer (VNA) is either directly sent to the device to excite the mechanical mode or used to generate modulated optical pump with a single-sideband modulator (SSBM), which creates a weak sideband at modulated frequency together with the optical pump. The receiver port of the VNA can be configured to measure the RF response of the device or the beating signal between the pump and sideband from the HSPD. This gives 4 different $S_{21}$ measurement schemes that provide a complete characterization of device properties. The results are presented in Fig.~\ref{fig3}b and are discussed in the following.

The electromechanical response ($S_\mathrm{ee}$) is first characterized. A sharp dip at $\omega_m=2\pi\times7.512$\,GHz can be seen corresponding to the mechanical resonance, whose frequency matches well with simulation results. Using a Lorentzian fitting, a linewidth of $\gamma=2\pi\times225$\,kHz can be extracted, corresponding to a high mechanical Q of 33\,k. The intrinsic loss rate and external coupling rate can also be fitted to be $\gamma_i=2\pi\times202$\,kHz and $\gamma_e=2\pi\times23$\,kHz, respectively. This allows us to calculate the intracavity phonon number in the mechanical resonator to be
\begin{equation}
    n_m=\frac{4\gamma_e}{\gamma^2}\times\frac{P_\mathrm{RF}}{\hslash\omega_m},
    \label{eq1}
\end{equation}
with $P_\mathrm{RF}$ being the RF drive power sent to the device.

\begin{figure*}[t]
\centering
\includegraphics[width=0.95\textwidth]{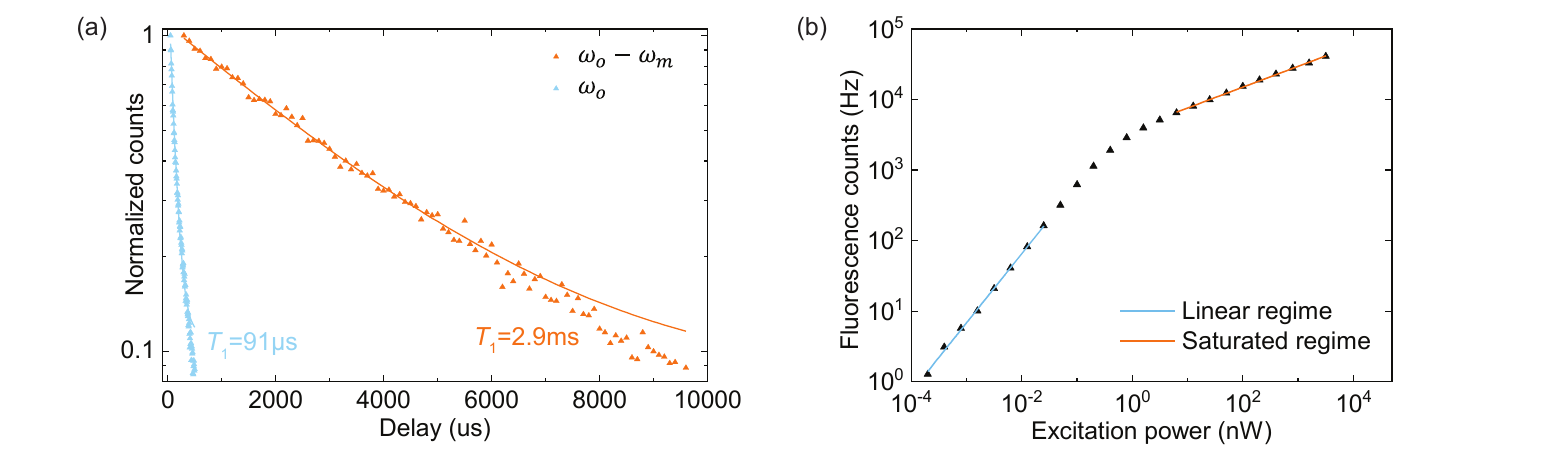}% Here is how to import EPS art
\caption{(a) Fluorescence decay of on-resonance ions at $\omega_o$ (blue) and off-resonance ions at $\omega_o-\omega_m$ (orange). The lines are exponential fitting with lifetime of 91\,{\textmu}s and 2.9\,ms, respectively. An over 30-fold decrease in lifetime can be achieved in the cavity due to the Purcell enhancement. (b) Fluorescence intensity ($I_f$) with varying excitation power ($I_o$). The emission will be saturated at higher power, providing a filtering of strong pump light. The lines corresponding to a fitting of $I_f\propto I_o^x$ with $x=1$ (blue) and $x=0.3$ (orange).}
\label{fig4}
\end{figure*}

By sending the RF drive instead to the SSBM and demodulate the beating signal from the HSPD, the reflection spectrum of the sideband ($S_\mathrm{oo}$) is extracted. This maps the optical resonance response of the device, showing a dip at $\omega_m$ due to the detuned drive. Similarly, the optical Q can be fitted to be 490\,k ($\kappa=2\pi\times400$\,MHz) with a intrinsic linewidth $\kappa_i=2\pi\times282$\,MHz and coupling linewidth $\kappa_e=2\pi\times118$\,MHz.

Finally, the bidirectional phonon-to-photon conversion can be demonstrated by measuring $S_\mathrm{eo}$ and $S_\mathrm{oe}$. This allows us to calibrate out the off-chip loss of each signal path and arrives at the phonon-mediated microwave-to-optical conversion efficiency \cite{andrews2014bidirectional}
\begin{equation}
    \eta=\sqrt{\frac{S_\mathrm{eo,p}S_\mathrm{oe,p}}{S_\mathrm{ee,bg}S_\mathrm{oo,bg}}},
    \label{eq2}
\end{equation}
where $S_\mathrm{eo,p},S_\mathrm{oe,p}$ are the $S_{21}$ at the peaks (on-resonance) of the conversion spectra and $S_\mathrm{ee,bg},S_\mathrm{oo,bg}$ are the $S_{21}$ at the background (off-resonance) of the corresponding spectra. With an on-chip optical power of $P_o=-12$\,dBm, we get $\eta=7.0\times10^{-5}$. On the other hand, the conversion efficiency can also be theoretically described by \cite{jiang2020efficient}
\begin{equation}
    \eta=\frac{\gamma_e}{\gamma}\frac{\kappa_e}{\kappa}\frac{4C}{(1+C)^2}.
    \label{eq3}
\end{equation}
Here, $C=4n_og_0^2/\kappa\gamma$ is the conversion cooperativity with $n_o$ being the intracavity pump photon number. From this the single-photon optomechanical coupling rate for our device can be calculated to be $g_0=2\pi\times3.2$\,kHz.

Using this number, we can compare the on-chip sideband photon count rate per intracavity phonon \cite{cohen2015phonon}
\begin{equation}
    \Gamma_\mathrm{sb}=\frac{\kappa_e}{\kappa}\frac{4n_og_0^2}{\kappa}
\end{equation}
with the photon count rate from the pump laser $\Gamma_p=P_o/\hslash\omega_o$ and get $\Gamma_\mathrm{sb}/\Gamma_p=6.4\times10^{-14}$. In other words, around 132\,dB extinction is needed from the filtering to achieve 1:1 signal-to-noise ratio (SNR) in direct sideband counting of a single intracavity phonon.

We now proceed to characterize the Er emission from our device with the single photon counting. Here, the pump laser is chopped into excitation pulses via an acousto-optic modulator (AOM) and the fluorescence emission from the device is measured by gating the SNSPD to bypass the excitation pulses and only collect the subsequent fluorescence. The Purcell enhancement of the cavity is first characterized by extracting the fluorescence lifetime. The results are shown in Fig.~\ref{fig4}a. When the excitation is at off-resonance frequency, the ions exhibit a lifetime of $T_{1,\mathrm{off}}$=2.9\,ms. However, when pumping on-resonance at the cavity frequency, the lifetime is considerably shortened to $T_{1,\mathrm{on}}$=91\,{\textmu}s, corresponding to a Purcell factor of 30. In counting the fluorescence from optomechanical sideband, since the emission rate of pump fluorescence (at $\omega_p=\omega_o-\omega_m$) is much lower, a short collection window can be used to only collect the sideband fluorescence (at $\omega_o$) via time-domain filtering. We also note that the cavity profile naturally provide a filter function to couple more sideband fluorescence to the waveguide, which further increases the extinction ratio between sideband and pump fluorescence.

\begin{figure*}[t]
\centering
\includegraphics[width=0.95\textwidth]{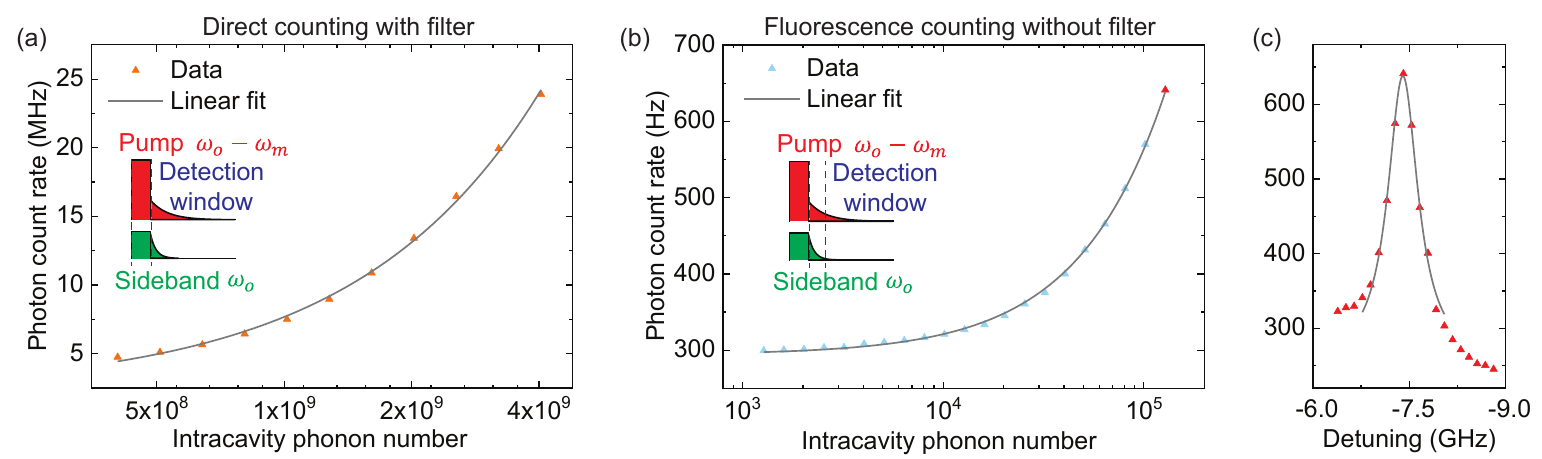}% Here is how to import EPS art
\caption{(a) Direct counting of the optomechanical sideband with an external filter. The inset shows the measurement scheme in which the detection window of the SNSPD aligns with the pump pulse. The length of the pump pulse is 50\,{\textmu}s with a duty cycle of 1\,\%. The photon count rate increases linear with intracavity phonon number, with a background count rate of 2.25\,MHz and per phonon count rate of 5.43\,mHz. (b) Instead of directly counting the sideband, the Er fluorescence emission can be counted in a 50\,{\textmu}s window after the pump pulse. By doing so, a much higher counting sensitivity can be achieved. Using a linear fit, background fluorescence count of 294\,Hz and per phonon count rate of 2.70\,mHz can be extracted. The equivalent filtering effect from the fluorescence counting can be calculated by comparing with the direct counting case. The red point corresponds to the measurement setting used in (c). (c) Change in fluorescence count when varying the pump detuning. The maximal phonon-to-photon conversion efficiency is achieved when the detuning is -$\omega_m$. The gray line is a Lorentzian fit.}
\label{fig5}
\end{figure*}

Apart from the filtering from cavity Purcell enhancement, another great advantage in counting the fluorescence lies in the saturable absorption of Er dopants. To demonstrate this effect, we fix the pump frequency at $\omega_o$ and vary the excitation power to measure the change in fluorescence intensity. The results are shown in Fig.~\ref{fig4}. The fluorescence intensity ($I_f$) under a certain excitation power ($I_o$) can be estimated by $I_f\propto I_o^x$. By fitting the data, we can see that at low power regime, the fluorescence intensity increase linearly with excitation power, i.e. $x=1$. When the excitation power is higher, however, the fluorescence emission will be saturated with $x\approx0.3$. In this way, the fluorescence phonon counting scheme provides a desirable filtering effect to suppress the strong pump while retaining the conversion efficiency of the weak sideband.

The phonon counting experiment is then performed by sending a continuous RF drive to generate intracavity phonons while applying a pulsed pump light at $\omega_p=\omega_o-\omega_m$ to convert the phonons to the optomechanical sideband. The on-chip pump power is set to be -11\,dBm, with a pulse length of 50\,{\textmu}s and a duty cycle of 1\,\% to minimize heating. To characterize the performance of the fluorescence counting method, we first consider the case of direct counting as a reference. An external filter (Alnair Lab BVF-300CL) is applied before the SNSPD to allow direct counting of the sideband photons. In other words, the detection window of the SNSPD is set to overlap with the pump pulse. The pump suppression ratio and insertion loss of the filter can be separated calibrated to be $\alpha_p=54$\,dB and $\alpha_\mathrm{sb}=7.0$\,dB, respectively. The measured results are shown in Fig.~\ref{fig5}a. The intracavity phonon number is calculated from Eq.~\ref{eq1} and varied by changing the applied RF power. Clearly, a linear increase in photon count rate can be seen when increasing the phonon number. Yet, a strong RF drive, i.e. large phonon numbers, is needed to get sufficient SNR in the counting. With a linear fit, we can extract the background count rate from the pump to be $\Gamma_p^{(1)}=2.25\pm0.13$\,MHz and the sideband count rate per intracavity phonon to be $\Gamma_\mathrm{sb}^{(1)}=5.43\pm0.07$\,mHz. Note that these numbers are the count rate directly read from the SNSPD, thus the system efficiency including the fiber loss and detector efficiency are all taken into account. 

We then turn to the filterless phonon counting scheme with Er fluorescence by simply collecting the photons emitted after the pump pulse. Note that the inhomogeneous broadening of Er transition (around 180\,GHz \cite{wang2022er}) is much larger than the mechanical frequency (7.5\,GHz) so that the spectral concentration of Er ions is similar for pump and sideband frequencies. The measurement results are plotted in Fig.~\ref{fig5}b. Here, the duration of the detection window is set to be 50\,{\textmu}s so that around 40\,\% of the sideband fluorescence can be collected. Setting the detection duration to be the same as pump pulse also enables straightforward comparison with the previous direct counting scheme. Similarly, using a linear fit we can get the background count rate from the pump to be $\Gamma_p^{(2)}=294\pm1$\,Hz and the count rate per intracavity phonon to be $\Gamma_\mathrm{sb}^{(2)}=2.70\pm0.15$\,mHz. We can also fix the RF and pump power and vary the pump wavelength to measure the change in fluorescence intensity, as illustrated in Fig.~\ref{fig5}c. The maximal sideband count rate is clearly achieved when the pump detuning is $-\omega_m$, which matches with theoretical prediction. The fluorescence count decreases toward the background value when the pump frequency is tuned away. A Lorentzian fitting of the trend would results in a FWHM matching the linewidth of the optical resonance.

The filtering effect from fluorescence measurement can then be calculated based on these results. Specifically, the equivalent pump suppression ratio of this process is $\Gamma_p^{(1)}/\Gamma_p^{(2)}\times\alpha_p=93$\,dB and the insertion loss for the sideband photons is $\Gamma_\mathrm{sb}^{(1)}/\Gamma_\mathrm{sb}^{(2)}\times\alpha_\mathrm{sb}=10$\,dB.

\section{Discussion and conclusion}
In conclusion, we have demonstrated filterless sideband phonon counting using fluorescence from a REI doped optomechanical cavity. Utilizing the smart-cut Er doped thin-film lithium niobate platform, we designed and fabricated a high-Q piezo-optomechanical Fabry-Perot resonator. With a complete characterization of microwave-to-optical conversion properties at 10\,mK, the mechanical Q, optical Q, and single-photon optomechanical coupling strength of our device are extracted to be 33\,k, 490\,k, and $2\pi\times$3.2\,kHz, respectively. We show that by counting the subsequent Er fluorescence instead of directly counting the sideband, an equivalent filtering of the pump can be achieved due to a combination of spectral hole burning and cavity Purcell effect. By comparing with the direct counting scheme, we show that the fluorescence detection scheme can realize a high pump suppression of 93\,dB and a sideband insertion loss of 10\,dB, corresponding to an equivalent extinction ratio of 83\,dB.

We note that the exact performance of the filtering would change with different pump power and collection window duration. When the pump power is higher, stronger hole burning effect would be present thus increase the pump suppression ratio. For the same reason, this measurement scheme would be most efficient for counting small number of phonon excitation, i.e. sideband fluorescence is not saturated, which complies with the quantum applications of mechanical resonators. Practically, however, when the pump power is too high, broadening of the homogeneous linewidth, thus the spectral hole linewidth, can be observed due to heating, which limits the filtering performance. By increasing the collection time $\tau$, the sideband fluorescence intensity increases as $1-e^{-\tau/T_{1,\mathrm{on}}}$, facilitating a higher counting efficiency. However, since the lifetime of pump fluorescence is much longer, its intensity will increase linearly for $\tau<<T_{1,\mathrm{off}}$, resulting in a decrease in SNR. There thus exists a trade-off in insertion loss and extinction ratio in our fluorescence filtering scheme. The exact measurement settings can then be optimized by tuning these parameters according to specific needs.

Despite of the high extinction ratio achieved, for the current device the background count from the pump fluorescence still correspond to around $10^5$ intracavity phonons and there is an apparent gap toward filterless counting of single phonons. The major limitation is the relatively low optomechanical coupling rate $g_0$. To increase $g_0$, devices with smaller mode volume can be studied. For example, by reducing the cavity length from 100\,{\textmu}m to 10\,{\textmu}m, the sideband photon count rate per phonon can be increased by 10 times. As the cavity Purcell enhancement is inversely proportional to mode volume, shorter fluorescence collection window can be used for shorter devices to reduce the background pump fluorescence. The reduced number of ions in the shortened cavity would also result in weaker pump fluorescence. These combined effect will result in a two-order-of-magnitude increase in counting sensitivity if the optical Q can be maintained after shortening the device. Alternatively, phononic crystal structures can also be explored to increase the mechanical Q as well as the optomechanical coupling. In the REI dopants point of view, several improvements can be made to further reduce the background fluorescence from the strong pump light. Firstly, an external magnetic field can be applied to the device to enable persistent spectral hole burning to other spin states. Also, due to the uniform doping in our device, the fluorescence from the coupling waveguide could also act as background noise source, which can be overcome by selective doping in the cavity. Applying these modifications would boost the sensitivity of our measurement approach toward single-phonon level.

Compared to using external filters in phonon counting, our unique proposal of embedding REI dopants directly in the optomechanical device offers multiple advantages, including better stability, reduced insertion loss, and simplified measurement configurations. Experiments such as mechanical mode thermometry and phonon addition can be readily envisioned. Apart from these purposes, our device configuration can also create new opportunities in quantum application of mechanical devices, including developing solid-state quantum memories for phonons \cite{wang2017quantum} and studying coupling between rare earth spins and mechanical motion \cite{ohta2024observation}.

\begin{acknowledgement}
The authors would like to thank Dr. Yong Sun, Dr. Lauren McCabe, Kelly Woods, and Dr. Michael Rooks for their assistance provided in the device fabrication. The fabrication of the devices was done at the Yale School of Engineering \& Applied Science (SEAS) Cleanroom and the Yale Institute for Nanoscience and Quantum Engineering (YINQE).
\end{acknowledgement}

\begin{funding}
This work was supported by the DOE Office of Science, the National Quantum Information Science Research Center, Codesign Center for Quantum Advantage (C2QA), Contract No. DESC0012704. HXT and JCX acknowledge funding support from the Air Force Office of Sponsored Research (AFOSR No. MURI FA9550-23-1-0338).
\end{funding}

\begin{authorcontributions}
H.X.T. and  L.Y. conceived the experiment. L.Y. designed and fabricated the device. L.Y. and J.X. contributed to the preparation of the device. L.Y. performed the experiment. L.Y. wrote the manuscript with contribution from all authors. H.X.T. supervised the work. All authors have accepted responsibility for the entire content of this manuscript and approved its submission.

\end{authorcontributions}

\begin{conflictofinterest}
Authors state no conflict of interest.
\end{conflictofinterest}

\begin{dataavailabilitystatement}
The datasets generated during the current study are available from the corresponding author on reasonable request.
\end{dataavailabilitystatement}

\bibliographystyle{ieeetr}
\bibliography{References}

\end{document}